\pdfoutput=1 
\documentclass[11pt]{article}
\usepackage[a4paper,margin=2.5cm]{geometry}
\usepackage{amsmath,amssymb,amsfonts}
\usepackage{bm}
\usepackage{graphicx}
\usepackage{booktabs}
\usepackage{placeins}
\usepackage{microtype}
\usepackage[round,sort]{natbib}
\usepackage[T1]{fontenc}
\usepackage{lmodern}
\usepackage[colorlinks=true,linkcolor=blue,citecolor=blue,urlcolor=blue]{hyperref}

\newcommand{\Pl}{\partial}
\newcommand{\ts}{\textstyle}
\newcommand{\bee}{\begin{equation}}
\newcommand{\ene}{\end{equation}}
\newcommand{\beea}{\begin{eqnarray}}
\newcommand{\enea}{\end{eqnarray}}
\newcommand{\non}{\nonumber}
\newcommand{\bg}{\boldsymbol{g}}
\newcommand{\fder}[2]{\frac{{\ts d #1}}{{\ts d #2}}}
\newcommand{\fpar}[2]{\frac{{\ts \Pl #1}}{{\ts \Pl #2}}}
\newcommand{\nder}[3]{\frac{{\ts d^{#1} #2}}{{\ts d #3^{#1}}}}
\newcommand{\sech}{\operatorname{sech}}
\newcommand{\artanh}{\operatorname{artanh}}
\newcommand{\peff}{p_{\rm eff}}
\newcommand{\ceff}{c_{\rm eff}}

\begin{document}

\begin{center}
{\Large\bf Self-gravitating isothermal equilibria in Eddington-inspired
Born--Infeld gravity: exact planar sheets and the Bonnor--Ebert critical
mass\par}
\vspace{1.2em}
{\large Nikhil Chakrabarti\par}
\vspace{0.6em}
Centre for Plasma Studies, Faculty of Science, Jadavpur University,\\
Kolkata 700032, India\\[0.3em]
Corresponding author: \texttt{nikhil.chakrabarti@gmail.com}
\end{center}

\begin{abstract}
We study self-gravitating fluid equilibria in the Newtonian limit of
Eddington-inspired Born--Infeld (EiBI) gravity, in which the Poisson
equation acquires a term proportional to the Laplacian of the density. We
show that this term is equivalent, for arbitrary flows, to an effective
pressure $p+\chi\rho^2/8$, so that equilibrium and radial stability reduce
exactly to Newtonian problems with a modified equation of state. For
pressure-bounded isothermal spheres we compute the EiBI generalization of
the Bonnor--Ebert critical mass as a function of the single parameter
$\varepsilon=\chi P_{\rm ext}/4c_s^4$, and confirm each change of stability
with a radial-pulsation analysis. Near $\varepsilon=0$ the critical mass
grows as $M_{\rm BE}(1+5.74\,\varepsilon)$. For $\varepsilon>0.0302$ no
isothermal sphere is radially unstable, so a positive EiBI parameter
removes the Bonnor--Ebert instability; for smaller positive $\varepsilon$
a stable high-density branch supported by the EiBI pressure follows the
unstable one. For $\varepsilon<-0.154$ the mass limit is set instead by
the vanishing of the central effective sound speed, and is a loss of
equilibrium rather than an instability. In planar geometry the same
reduction yields an exact first integral, closed-form isothermal sheets
with a maximum column density $2c_s^2/\sqrt{\pi G|\chi|}$ for $\chi<0$,
and a complete classification of adiabatic $\gamma=3$ sheets.
\end{abstract}


\section{Introduction}
The isothermal self-gravitating sphere bounded by an external pressure is
one of the central models of gravitational fluid dynamics. \citet{Bonnor1956} and
\citet{Ebert1955} showed that such a sphere has a maximum
mass, $M_{\rm BE}=1.182\,c_s^4/(G^{3/2}P_{\rm ext}^{1/2})$, above which no
stable equilibrium exists, and this critical mass remains the standard
criterion for the gravitational collapse of molecular cloud
cores \citep{Alves2001}. Its planar counterpart is the isothermal sheet of
\citet{Spitzer1942},
$\rho=\rho_0\sech^2(x/\sqrt{2}\lambda_J)$, the standard model for the
vertical structure of gaseous discs \citep{BinneyTremaine2008} and, since
\citet{Ledoux1951}, a reference state for gravitational instability
in stratified media. Self-gravitating media are also studied extensively
in plasma physics, for example in dusty, quantum and viscoelastic
plasmas \citep{Janaki2011,AkbariMoghanjoughi2014,Rozina2017,ShuklaAvinash2019}.

In Eddington-inspired Born--Infeld (EiBI)
gravity \citep{Banados2010,BeltranJimenez2018} the field equations reduce
to those of general relativity in vacuum but differ inside matter. In the
non-relativistic limit the Poisson equation becomes \citep{Banados2010}
\bee
\nabla^2\psi=4\pi G\rho+\frac{\chi}{4}\nabla^2\rho ,
\label{eq:poisson_intro}
\ene
where $\chi$ is the Born--Infeld parameter. EiBI gravity has been applied
to neutron and compact stars \citep{Pani2011,Pani2012,Harko2013}, low-mass
stars \citep{Guerrero2022}, and the Sun \citep{Das2024Solar}; reviews are
given by \citet{BeltranJimenez2018} and \citet{Olmo2020}, and constraints on
$\chi$ come from astrophysics and cosmology \citep{Avelino2012} and, more
recently, hadron structure \citep{Goharipour2025}. In the Newtonian limit,
Eq.~(\ref{eq:poisson_intro}) has been used to study the Jeans instability
in kinetic theory \citep{DeMartino2017}, with quantum
corrections \citep{Bessiri2021}, in viscoelastic and polytropic
astrofluids \citep{Yang2023,Hazarika2024}, and in the
nonlinear regime \citep{DasKarmakar2024Puls,DasKarmakar2024NL}. These
analyses perturb a homogeneous background and find that $\chi>0$
stabilizes and $\chi<0$ destabilizes the medium. How EiBI gravity changes
the critical mass of a bounded, self-consistently stratified cloud, the
quantity that actually decides collapse, has not been determined.

The key simplification, noted by \citet{Pani2011} for hydrostatic equilibrium, is that the $\chi$
term can be absorbed into the equation of state through the effective
pressure
\bee
\peff=p+\frac{\chi}{8}\rho^2 .
\label{eq:peff_intro}
\ene
EiBI gravity therefore acts, inside matter, as an additional $\gamma=2$
polytropic pressure, positive for $\chi>0$ and negative for $\chi<0$. We
show in Sec.~\ref{sec:sphere} that this equivalence holds for arbitrary
time-dependent flows, so that the Newtonian theory of radial stability
applies without modification. The same quadratic term appears in the
Thomas--Fermi description of Bose--Einstein-condensed dark matter with
contact interactions \citep{Boehmer2007,Chavanis2011}, so the results also
apply, with a change of interpretation, to such fluids.

In this paper we use Eq.~(\ref{eq:peff_intro}) to treat both geometries.
The planar problem admits exact solutions and isolates the role of the
effective sound speed; the spherical problem gives the EiBI Bonnor--Ebert
mass. Our main results are as follows.
\begin{enumerate}
\item A general planar first integral,
$\peff(0)-\peff(x)=2\pi G\Sigma^2(x)$, relating the effective pressure to
the column density $\Sigma$, with closed-form isothermal sheets for all
$\hat\chi\equiv\chi\rho_0/4c_s^2>-1$ and a maximum column density for
$\chi<0$ (Sec.~\ref{sec:iso}).
\item For $\gamma=3$ sheets, a proof that negative-$\chi$ equilibria
terminate where the effective sound speed vanishes, compact
positive-$\chi$ sheets with a finite edge slope, and two elementary
solutions (Sec.~\ref{sec:adia}).
\item The EiBI Bonnor--Ebert critical mass $\mu_{\max}(\varepsilon)$ as a
function of $\varepsilon=\chi P_{\rm ext}/4c_s^4$, with three regimes: a
shifted turning point for $\varepsilon_*<\varepsilon<\varepsilon_c$, no
instability at all for $\varepsilon>\varepsilon_c\simeq0.0302$, and a mass
limit set by vanishing central sound speed for
$\varepsilon<\varepsilon_*\simeq-0.154$ (Sec.~\ref{sec:sphere}).
\end{enumerate}

The paper is organized as follows. Section~\ref{sec:model} introduces the
effective barotropic formulation and the planar first integral.
Sections~\ref{sec:iso} and~\ref{sec:adia} treat isothermal and adiabatic
($\gamma=3$) sheets. Section~\ref{sec:sphere} treats pressure-bounded
isothermal spheres and their radial stability. Section~\ref{sec:disc}
discusses the results and Sec.~\ref{sec:concl} concludes.

\section{Model and effective barotropic formulation}
\label{sec:model}
\subsection{Governing equations}
We consider a compressible self-gravitating fluid in the Newtonian limit of
EiBI gravity,
\bee
\fpar{\rho}{t}+\nabla\cdot(\rho\mathbf{v})=0,\qquad
\rho\left(\fpar{\mathbf{v}}{t}+\mathbf{v}\cdot\nabla\mathbf{v}\right)
=-\nabla p+\rho\bg ,
\label{eq:fluid}
\ene
with $\bg=-\nabla\psi$ and $\psi$ determined by
Eq.~(\ref{eq:poisson_intro}). The additional term couples the potential
directly to the local density, so it matters wherever the density varies
on short scales.

\subsection{Planar equilibrium and normalization}
For a static ($\mathbf{v}=0$) configuration that depends only on $x$ and is
symmetric about $x=0$, Eqs.~(\ref{eq:fluid}) and (\ref{eq:poisson_intro})
reduce to
\bee
\fder{p}{x}+\rho\fder{\psi}{x}=0,
\qquad
\nder{2}{\psi}{x}=4\pi G\rho+\frac{\chi}{4}\nder{2}{\rho}{x}.
\label{eq:planar}
\ene
We consider the isothermal law $p=c_s^2\rho$ and the polytropic law
$p=K\rho^\gamma$. For the polytropic case, $c_s^2\equiv\gamma K\rho_0^{\gamma-1}$
denotes the sound speed at the central density $\rho_0$. The value
$\gamma=3$, treated in Section~\ref{sec:adia}, is the adiabatic index for
one-dimensional compression of a gas with a single translational degree of
freedom. It is familiar in plasma physics from fluid closures of
one-dimensional kinetic models. For this value the equilibrium problem
becomes polynomial.

We introduce dimensionless variables
\bee
\hat\rho=\frac{\rho}{\rho_0},\quad
\hat x=\frac{x}{\lambda_J},\quad \hat\psi=\frac{\psi}{c_s^2},\quad
\hat p=\frac{p}{\rho_0c_s^2},\quad \hat\chi=\frac{\chi\rho_0}{4c_s^2},
\label{eq:norm}
\ene
where $\lambda_J=(c_s^2/4\pi G\rho_0)^{1/2}$ is the Jeans length. The
parameter $\hat\chi$ measures the EiBI pressure $\chi\rho_0^2/8$ against
the thermal pressure at the center. Hats are dropped from now on. In these
units Eq.~(\ref{eq:planar}) becomes
\bee
\fder{p}{x}=-\rho\fder{\psi}{x},\qquad
\nder{2}{\psi}{x}=\rho+\chi\nder{2}{\rho}{x},
\label{eq:planar_dl}
\ene
with boundary conditions $\rho(0)=1$ and $\rho'(0)=0$.

\subsection{Effective pressure and the planar first integral}
We split the potential as $\psi=\psi_N+\chi\rho$, where $\psi_N''=\rho$ is
the Newtonian potential. Hydrostatic balance then becomes
\bee
\fder{\peff}{x}=-\rho\fder{\psi_N}{x},\qquad
\peff(\rho)=p(\rho)+\frac{\chi}{2}\rho^2,
\label{eq:peff}
\ene
which is the dimensionless form of Eq.~(\ref{eq:peff_intro}). We define the
effective enthalpy $h(\rho)=\int d\peff/\rho$ and the effective sound speed
\bee
\ceff^2(\rho)=\fder{\peff}{\rho}=\fder{p}{\rho}+\chi\rho .
\label{eq:ceff}
\ene
The equilibrium then satisfies
\bee
\nder{2}{h}{x}=-\rho .
\label{eq:hpp}
\ene
This is the Newtonian equation for a fluid with pressure $\peff$.

Now let $\Sigma(x)=\int_0^x\rho\,dx'$ be the column density measured from
the midplane. Since $\psi_N'(0)=0$, Gauss's law gives $\psi_N'=\Sigma$,
so the Newtonian part of the field is $g_N=-\Sigma$ for $x>0$. Using
$\rho=\Sigma'$ in Eq.~(\ref{eq:peff}) gives
$d\peff/dx=-\Sigma\,d\Sigma/dx$, and integrating from the midplane,
\bee
\peff(1)-\peff(\rho)=\frac{1}{2}\Sigma^2 .
\label{eq:firstint}
\ene
In dimensional form this reads $\peff(0)-\peff(x)=2\pi G\Sigma^2$.
Equation~(\ref{eq:firstint}) holds for any barotropic equation of state. It
is the first integral of Eq.~(\ref{eq:hpp}), since $h'=-\Sigma$. It gives
the integration constant a physical meaning: the effective midplane
pressure supports the column above it. In particular, a sheet whose
density vanishes at its boundary has half-column
\bee
\Sigma_b=\sqrt{2\peff(1)} .
\label{eq:Sigmab}
\ene

The field that acts on the fluid is $g=-\psi'=g_N-\chi\rho'$. Using
$h'=\ceff^2\rho'/\rho=-\Sigma$ gives
\bee
g=-\frac{\Sigma}{\ceff^2}\fder{p}{\rho}.
\label{eq:g_general}
\ene
Outside the matter, $\rho=0$ and $g=g_N=\mp\Sigma_b$, which is the Newtonian
exterior field. At the midplane, Eq.~(\ref{eq:hpp}) gives
$\ceff^2(1)\rho''(0)=-1$, so
\bee
\rho''(0)=-\frac{1}{\ceff^2(1)} .
\label{eq:curv}
\ene
A regular density maximum at $x=0$ therefore requires $\ceff^2(1)>0$. For
both equations of state considered below, this condition is
$\chi>-1$.

\subsection{Linear response and local Jeans criterion}
The effective sound speed controls the linear response in the same way. We
linearize Eqs.~(\ref{eq:fluid}) and (\ref{eq:poisson_intro}) about a
uniform state, with the usual Jeans swindle. In dimensional form this
gives
\bee
\omega^2=\left(c_s^2+\frac{\chi}{4}\rho_0\right)k^2-4\pi G\rho_0
=\ceff^2k^2-4\pi G\rho_0 ,
\label{eq:disp}
\ene
so the Jeans wavenumber is $k_J^2=4\pi G\rho_0/\ceff^2$. A positive $\chi$
stiffens the medium and a negative $\chi$ softens it, in agreement with
the stability trends reported for EiBI astrofluids \citep{Yang2023,Hazarika2024}.
A kinetic treatment leads to the same modified Jeans
scale \citep{DeMartino2017}. Where $\ceff^2\to0$ the local Jeans length
vanishes. Where $\ceff^2<0$, perturbations of every wavelength grow even
without self-gravity. The situation is analogous to the way quantum and
degeneracy pressures modify the Jeans criterion in quantum
plasmas \citep{AkbariMoghanjoughi2014}. As shown below, the surfaces on
which $\ceff^2=0$ control the existence and termination of the nonlinear
equilibria.

\section{Isothermal sheets}
\label{sec:iso}
\subsection{First integral}
For $p=\rho$ we have
\bee
\peff=\rho+\frac{\chi}{2}\rho^2,\qquad \ceff^2=1+\chi\rho,\qquad
h=\ln\rho+\chi\rho .
\ene
Since hydrostatic balance gives $\psi=-\ln\rho$, the equilibrium equation
(\ref{eq:hpp}) reads
\bee
\nder{2}{\ln\rho}{x}=-\rho-\chi\nder{2}{\rho}{x}.
\label{eq:iso_ode}
\ene
The first integral (\ref{eq:firstint}), together with
$\Sigma=-h'=-(1+\chi\rho)\rho'/\rho$, gives
\bee
\fder{\rho}{x}=\mp\frac{\rho\sqrt{(1-\rho)\,[2+\chi(1+\rho)]}}{1+\chi\rho}.
\label{eq:iso_first}
\ene
This agrees with the result of direct integration by parts of
Eq.~(\ref{eq:iso_ode}). For $\chi>-1$, both $1+\chi\rho$ and
$S(\rho)=2+\chi(1+\rho)$ are positive on $0<\rho\le1$. The right-hand side
is therefore regular and vanishes only at the midplane. By
Eq.~(\ref{eq:curv}), the central expansion is
\bee
\rho(x)\simeq1-\frac{x^2}{2(1+\chi)},\qquad |x|\ll1 .
\label{eq:iso_center}
\ene
The curvature diverges as $\chi\to-1^+$, when $\ceff^2(1)=1+\chi\to0$. At
$\chi=-1$ both the numerator and denominator of Eq.~(\ref{eq:iso_first})
vanish at $\rho=1$, and no regular maximum exists. Physically, the central
effective compressibility changes sign there.

\subsection{Newtonian limit}
\begin{figure}[t]
\centering
\includegraphics[width=\linewidth]{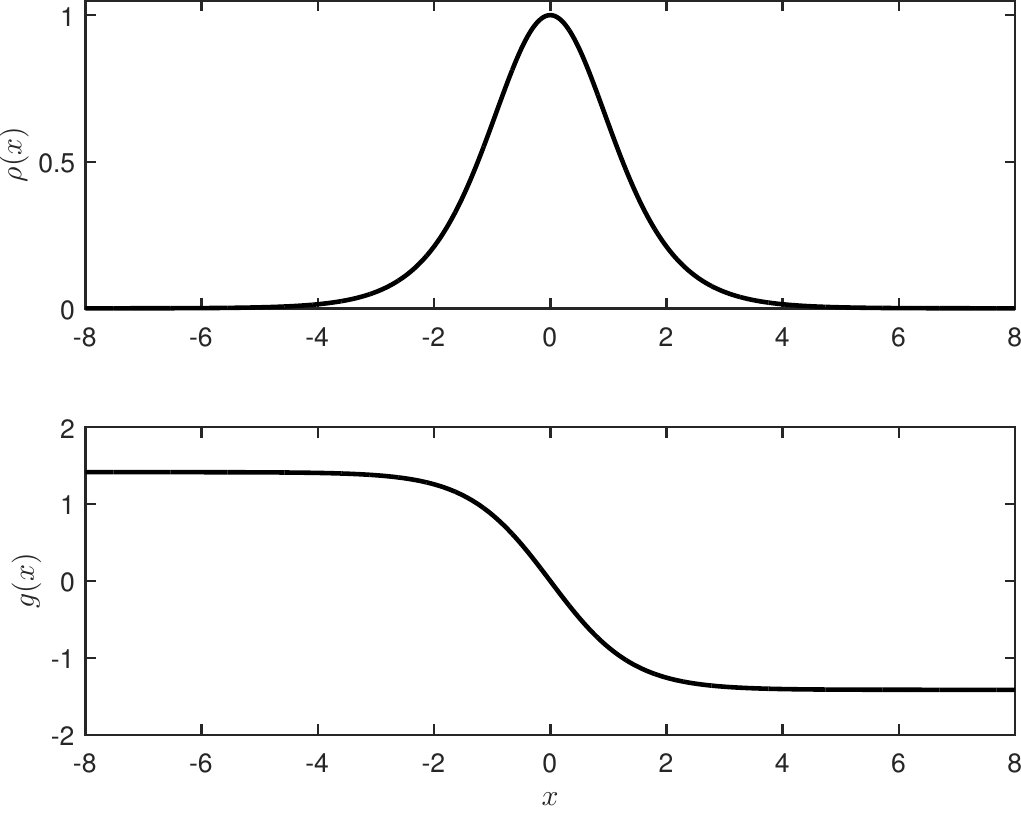}
\caption{Density $\rho(x)$ (top) and gravitational field $g(x)$ (bottom) of
the isothermal sheet for $\chi=0$, i.e. the Spitzer
$\mathrm{sech}^2$ profile of Eq.~(\ref{eq:spitzer}). The field vanishes at
the midplane and tends to $\mp\sqrt2$, the Newtonian value $\mp\Sigma_b$.}
\label{fig:zero_chi}
\end{figure}

For $\chi=0$, Eq.~(\ref{eq:iso_first}) integrates to Spitzer's
solution \citep{Spitzer1942},
\bee
\rho=\rho_0\sech^2\!\left(\frac{x}{\sqrt2\,\lambda_J}\right),\qquad
g=-\frac{\sqrt2\,c_s^2}{\lambda_J}\tanh\!\left(\frac{x}{\sqrt2\,\lambda_J}\right),
\label{eq:spitzer}
\ene
with dimensions restored. This profile is shown in
Fig.~\ref{fig:zero_chi} and serves as the reference configuration.

\subsection{Exact solutions for $\chi\neq0$}
Separating variables in Eq.~(\ref{eq:iso_first}) gives
\bee
|x|=\int_\rho^1\frac{(1+\chi s)\,ds}{s\sqrt{(1-s)\,[2+\chi(1+s)]}} .
\label{eq:iso_quad}
\ene
The integral splits into a logarithmic part and an elementary
part. For $\chi>0$,
\beea
\frac{|x|}{\sqrt2}&=&\sqrt{\frac{2}{2+\chi}}\;
\artanh\!\left(\sqrt{\frac{(2+\chi)(1-\rho)}{2+\chi(1+\rho)}}\right)\non\\
&&+\sqrt{2\chi}\;\arctan\!\left(\sqrt{\frac{\chi(1-\rho)}{2+\chi(1+\rho)}}\right),
\label{eq:iso_pos}
\enea
and for $-1<\chi<0$,
\beea
\frac{|x|}{\sqrt2}&=&\sqrt{\frac{2}{2+\chi}}\;
\artanh\!\left(\sqrt{\frac{(2+\chi)(1-\rho)}{2+\chi(1+\rho)}}\right)\non\\
&&-\sqrt{-2\chi}\;\artanh\!\left(\sqrt{\frac{-\chi(1-\rho)}{2+\chi(1+\rho)}}\right).
\label{eq:iso_neg}
\enea
The second term of Eq.~(\ref{eq:iso_neg}) enters with a negative sign. It
is the analytic continuation of the $\arctan$ term in
Eq.~(\ref{eq:iso_pos}), obtained using $\arctan(iz)=i\,\artanh z$. Its
argument is less than unity for $\chi>-1$. We have checked
Eqs.~(\ref{eq:iso_pos}) and (\ref{eq:iso_neg}) against direct numerical
quadrature of Eq.~(\ref{eq:iso_quad}). They agree to better than
$10^{-10}$ over $10^{-4}\le\rho<1$ for all values of $\chi$ used here.
Both expressions reduce to Eq.~(\ref{eq:spitzer}) as $\chi\to0$.

As $\rho\to0$ only the first $\artanh$ term diverges. The tail is therefore
exponential,
\bee
\rho\propto\exp\!\left(-\sqrt{2+\chi}\,|x|\right),\qquad |x|\to\infty ,
\label{eq:iso_tail}
\ene
and the sheet extends to infinity for every $\chi>-1$.

\subsection{Structure of the solutions}
\begin{figure}[t]
\centering
\includegraphics[width=\linewidth]{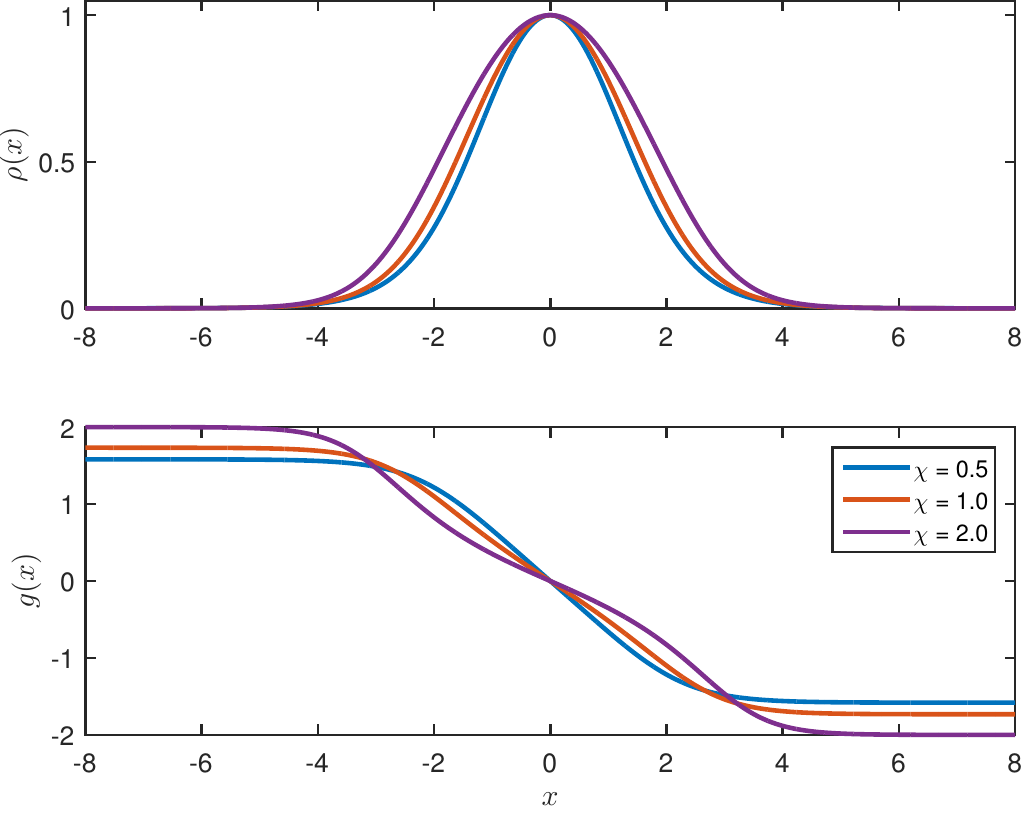}
\caption{Isothermal sheets for $\chi=0.5$, $1$ and $2$: density (top) and
gravitational field $g=\rho'/\rho$ (bottom). Increasing $\chi$ broadens the
core and raises the asymptotic field $\mp\sqrt{2+\chi}$. The field curves
cross because $|g|\simeq|x|/(1+\chi)$ near the midplane (see
Eq.~(\ref{eq:iso_g})).}
\label{fig:pos_chi}
\end{figure}

\begin{figure}[t]
\centering
\includegraphics[width=\linewidth]{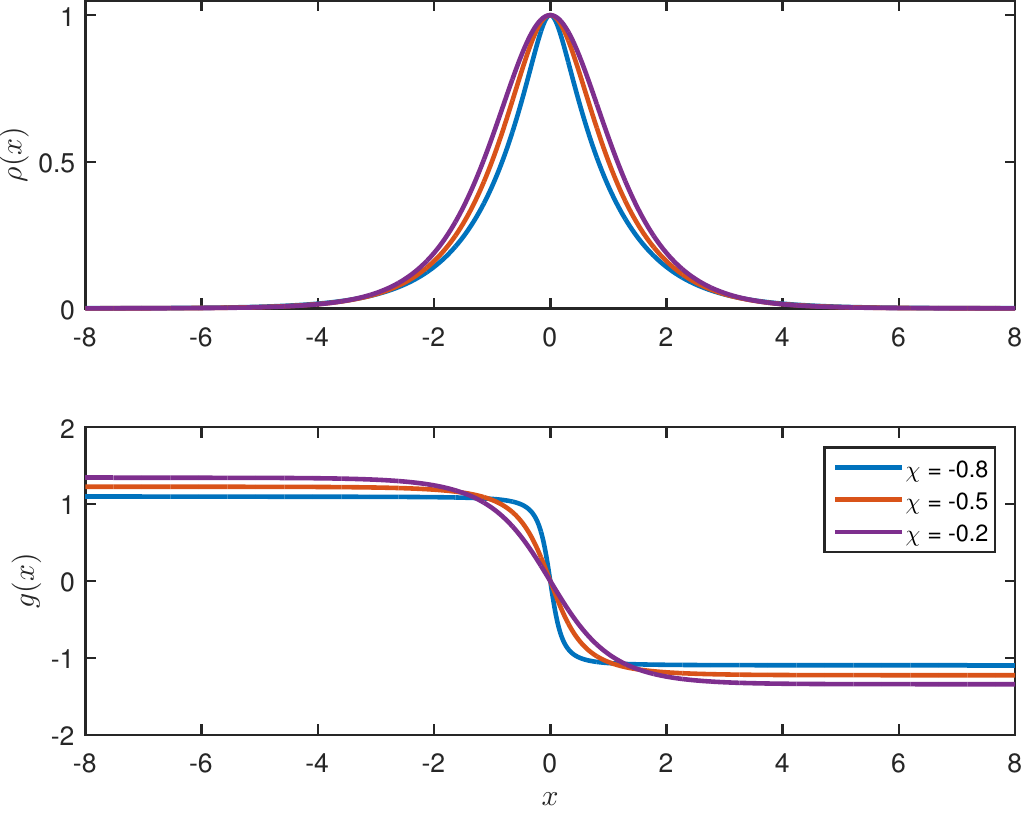}
\caption{Isothermal sheets for $\chi=-0.8$, $-0.5$ and $-0.2$: density
(top) and gravitational field (bottom). The profiles are regular, with a
sharper core as $\chi\to-1$ (central curvature $-1/(1+\chi)$). The
exponential tail $\propto e^{-\sqrt{2+\chi}|x|}$ is shallower, and the field
tends to $\mp\sqrt{2+\chi}$.}
\label{fig:neg_chi}
\end{figure}

Figures~\ref{fig:pos_chi} and \ref{fig:neg_chi} show the profiles for
positive and negative $\chi$. The profiles were obtained numerically from
Eq.~(\ref{eq:iso_first}). They agree with Eqs.~(\ref{eq:iso_pos}) and
(\ref{eq:iso_neg}). Because the first-order equation is degenerate at
$\rho=1$, where $\rho'=0$, the profiles can be generated either from the
quadrature (\ref{eq:iso_quad}) or by integrating Eq.~(\ref{eq:iso_ode})
from $\rho(0)=1$, $\rho'(0)=0$.

The EiBI parameter affects the core and the tail in opposite directions.
The half-width at half-maximum grows with $\chi$ (Table~\ref{tab:summary}),
because the central curvature $-1/(1+\chi)$ decreases in magnitude. The
decay rate $\sqrt{2+\chi}$ of the tail, however, \emph{increases} with $\chi$.
For $-1<\chi<0$ the core is sharper but the tail is more extended than in
the Newtonian case. For example, the density falls to $10^{-2}$ at
$|x|=4.42$ for $\chi=-0.8$, compared with $|x|=4.23$ for $\chi=0$.

The same competition explains the crossing of the field profiles in
Figs.~\ref{fig:pos_chi} and \ref{fig:neg_chi}. From
Eq.~(\ref{eq:g_general}),
\bee
g=\frac{1}{\rho}\fder{\rho}{x}=-\frac{\Sigma}{1+\chi\rho}.
\label{eq:iso_g}
\ene
Near the midplane $g\simeq-x/(1+\chi)$, whose magnitude decreases with
$\chi$. Far away, $g\to\mp\Sigma_b=\mp\sqrt{2+\chi}$, whose magnitude
increases with $\chi$. The curves for different $\chi$ must therefore
cross at intermediate $|x|$. This is a general property of the family,
not a feature of the particular values shown.

\subsection{Column density and the maximum column of negative-$\chi$ sheets}
Since $\rho\to0$ at infinity, Eq.~(\ref{eq:Sigmab}) gives the half-column
$\Sigma_b=\sqrt{2+\chi}$. In dimensional units the total surface density
$\Sigma_{\rm tot}=2\Sigma_b$ of the sheet satisfies
\bee
c_s^2\rho_0+\frac{\chi}{8}\rho_0^2=\frac{\pi G}{2}\,\Sigma_{\rm tot}^2 .
\label{eq:iso_column}
\ene
Equation~(\ref{eq:iso_column}) is the planar analogue of a mass--radius
relation. It gives the midplane density of a sheet of prescribed column.
For $\chi>0$ the EiBI pressure supports part of the column, so
$\rho_0$ is lower than in the Newtonian case. For $\chi<0$ the left-hand side
of Eq.~(\ref{eq:iso_column}) has a maximum, $2c_s^4/|\chi|$, at
$\rho_0=4c_s^2/|\chi|$, i.e. exactly at the critical value $\hat\chi=-1$.
Isothermal sheets in EiBI gravity with $\chi<0$ therefore exist only for
\bee
\Sigma_{\rm tot}\le\Sigma_{\max}=\frac{2c_s^2}{\sqrt{\pi G|\chi|}} .
\label{eq:Sigmamax}
\ene
A sheet with a larger column has no regular equilibrium. The critical case
$\hat\chi=-1$ is thus the equilibrium of maximum column. The Newtonian
sheet has no such limit, since $\Sigma_{\max}\to\infty$ as $\chi\to0^-$.
The branch $\rho_0>4c_s^2/|\chi|$, which would correspond to $\hat\chi<-1$,
is excluded because its central effective compressibility is negative.
Section~\ref{sec:sphere} shows that this end-point limit
persists in spherical geometry, where it competes with the Bonnor--Ebert
turning point. It shows that the regularity bound $\hat\chi>-1$
is a physical limit on the column that negative-$\chi$ gravity can support,
not merely a mathematical restriction.

\section{Adiabatic sheets with $\gamma=3$}
\label{sec:adia}
\subsection{First integral}
For $p=K\rho^3$ with $c_s^2=3K\rho_0^2$, the dimensionless pressure is
$p=\rho^3/3$. Hydrostatic balance gives $\psi=-\rho^2/2$, where the
constant is chosen so that $\psi\to0$ as $\rho\to0$. Then
\bee
\peff=\frac{\rho^3}{3}+\frac{\chi}{2}\rho^2,\quad
\ceff^2=\rho(\rho+\chi),\quad
h=\frac{\rho^2}{2}+\chi\rho ,
\label{eq:adia_eff}
\ene
and Eq.~(\ref{eq:hpp}) reads
\bee
\frac12\nder{2}{\rho^2}{x}=-\rho-\chi\nder{2}{\rho}{x}.
\label{eq:adia_ode}
\ene
The first integral (\ref{eq:firstint}) with $\Sigma=-(\rho+\chi)\rho'$ is
\bee
(\rho+\chi)\fder{\rho}{x}=\mp\sqrt{R(\rho)},\qquad
R(\rho)=\frac23(1-\rho^3)+\chi(1-\rho^2).
\label{eq:adia_first}
\ene
The factor multiplying $\rho'$ is $\ceff^2/\rho$. It vanishes where the
effective sound speed vanishes. Near the midplane, Eq.~(\ref{eq:curv}) again
gives $\rho\simeq1-x^2/[2(1+\chi)]$. This coincides with the isothermal
result because $\ceff^2(1)=1+\chi$ in both cases. A regular equilibrium
again requires $\chi>-1$.

\subsection{Newtonian limit}
For $\chi=0$, Eq.~(\ref{eq:adia_first}) integrates to
\bee
\frac12\,{}_2F_1\!\left(\tfrac12,\tfrac23;\tfrac53;1\right)
-\frac{\rho^2}{2}\,{}_2F_1\!\left(\tfrac12,\tfrac23;\tfrac53;\rho^3\right)
=\sqrt{\tfrac23}\,|x| .
\label{eq:adia_newton}
\ene
The density vanishes at
\bee
x_e=\frac12\sqrt{\frac32}\,
\frac{\Gamma(\tfrac53)\Gamma(\tfrac12)}{\Gamma(\tfrac76)}\simeq1.056,
\ene
where $\rho'\propto-1/\rho$ diverges, while $g=\rho\rho'=-\sqrt{2/3}$
remains finite and matches the exterior field $-\Sigma_b$.

\subsection{Root structure and the accessible density interval}
For $\chi\ne0$ we write
\bee
R(\rho)=\frac23(1-\rho)\,Q(\rho),\qquad
Q(\rho)=\rho^2+a\rho+a,
\label{eq:Q}
\ene
with $a=1+\frac32\chi$, so that the solution is given by the quadrature
\bee
|x|=\sqrt{\frac32}\int_\rho^1\frac{(s+\chi)\,ds}{\sqrt{(1-s)\,Q(s)}} .
\label{eq:adia_quad}
\ene
The roots of $Q$ are
$\rho_\pm=\frac12\left[-a\pm\sqrt{a(a-4)}\right]$. They are real for
$\chi\le-2/3$ or $\chi\ge2$ and complex for $-2/3<\chi<2$. For $\chi\ge2$ both
are negative. For $-1<\chi<-2/3$ one has $\rho_-<0<\rho_+<1$, since
$Q(0)=a<0$ and $Q(1)=3(1+\chi)>0$.

The positive root $\rho_+$ is never reached by the physical branch.
The vertex of $Q$ lies at $\rho=-a/2$, and $-a/2<-\chi$ whenever $\chi<2$,
so $Q$ is increasing on $[-\chi,1]$. Moreover,
\bee
Q(-\chi)=\frac12(2-\chi)(1+\chi)>0,\qquad -1<\chi<0 .
\ene
Hence $Q>0$, and therefore $R>0$, on the whole interval $-\chi\le\rho<1$,
and $\rho_+<-\chi$ whenever $\rho_+$ is real. The same argument with
$Q(0)=a>0$ shows that $R>0$ on $0\le\rho<1$ for every $\chi>0$. The
accessible density interval is therefore
\bee
\rho_b\le\rho\le1,\qquad
\rho_b=\begin{cases}0, & \chi\ge0,\\ -\chi, & -1<\chi<0,\end{cases}
\label{eq:rhob}
\ene
for every $\chi>-1$. On this interval the only zero of $R$ is the midplane
and the only zero of $\rho+\chi$ is the lower endpoint. The roots of the
cubic classify the elliptic form of Eq.~(\ref{eq:adia_quad}), but they have
no effect on the physical solution.

\subsection{Positive $\chi$: compact sheets with a finite edge slope}
For $\chi>0$ the density reaches zero at a finite distance $x_e$, listed in
Table~\ref{tab:summary}. There $R(0)=2/3+\chi$, so the edge slope is finite,
\bee
\left.\fder{\rho}{x}\right|_{x_e}=-\frac{\Sigma_b}{\chi},\qquad
\Sigma_b=\sqrt{\frac23+\chi},
\label{eq:edge}
\ene
and $\rho\simeq(\Sigma_b/\chi)(x_e-|x|)$ near the edge. The reason is that
the quadratic EiBI term dominates $\peff$ at low density, so the outer
layers behave as a $\gamma=2$ ($n=1$) polytrope, which has a linear edge.
This contrasts with the infinite slope of the Newtonian $\gamma=3$ edge.
The field acting on the fluid, $g=\rho\rho'$, therefore tends to zero at
the edge, while the exterior field is $g=\mp\Sigma_b$. The Newtonian part
$g_N=-\Sigma$ is continuous, and the jump in $g$ is carried entirely by the
term $-\chi\rho'$. This is the planar counterpart of the surface behavior
of EiBI stars discussed by \citet{PaniSotiriou2012} and \citet{Kim2014}.

For $\chi=2$ the roots of $Q$ merge at $\rho=-2$, so that
$Q=(\rho+2)^2$ cancels the numerator of Eq.~(\ref{eq:adia_quad}). The
solution is then elementary:
\bee
\rho=1-\frac{x^2}{6},\qquad g=-\frac{x}{3}\left(1-\frac{x^2}{6}\right),
\qquad |x|\le\sqrt6 .
\label{eq:chi2}
\ene
One can verify directly that Eq.~(\ref{eq:chi2}) satisfies
Eq.~(\ref{eq:adia_ode}). Its edge slope $-\sqrt6/3$ agrees with
Eq.~(\ref{eq:edge}). It provides a simple exact benchmark for numerical
solvers.

\begin{figure}[t]
\centering
\includegraphics[width=\linewidth]{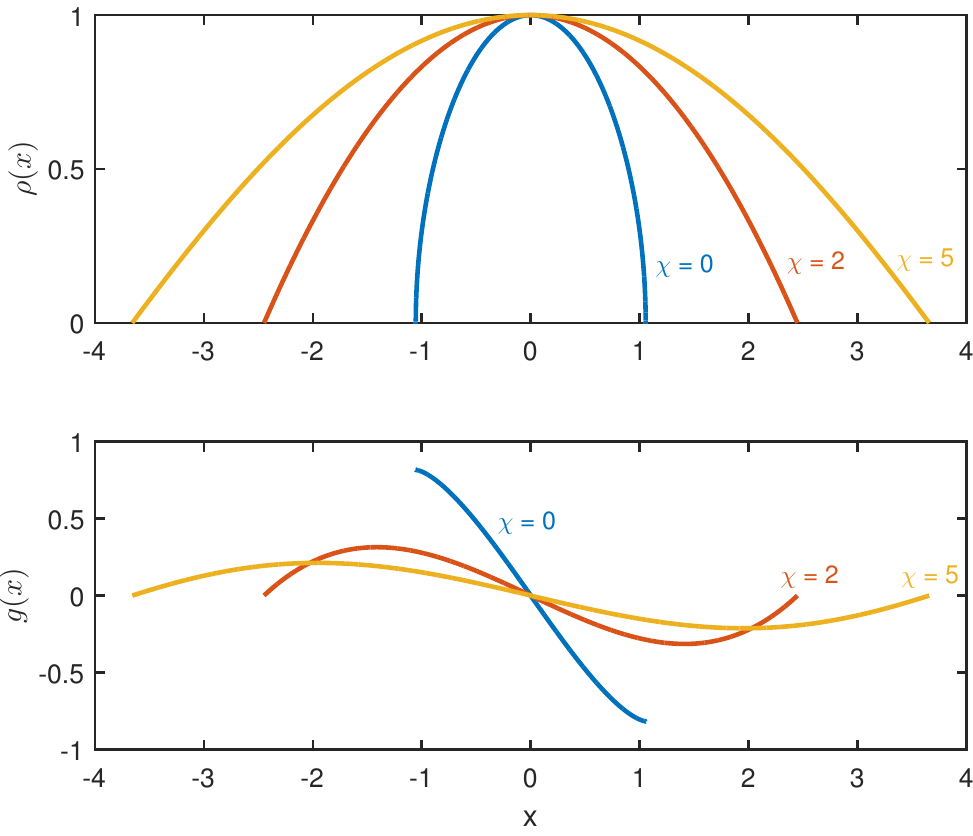}
\caption{Adiabatic ($\gamma=3$) sheets for $\chi=0$, $2$ and $5$: density
(top) and field $g=\rho\rho'$ (bottom). All three have compact support. For
$\chi=0$ the edge ($x_e\simeq1.056$) has an infinite slope. For $\chi>0$ the
EiBI pressure makes the edge slope finite, $-\Sigma_b/\chi$, and $g\to0$
at the edge. The $\chi=2$ curve is the exact parabola of
Eq.~(\ref{eq:chi2}), with $x_e=\sqrt6$.}
\label{fig:comparison_chi}
\end{figure}

\subsection{Negative $\chi$: termination at the zero of the effective sound speed}
For $-1<\chi<0$ the equilibrium ends at $\rho_b=-\chi$, where $\ceff^2=0$,
at the finite distance
\bee
x_s=\sqrt{\frac32}\int_{-\chi}^1\frac{(s+\chi)\,ds}{\sqrt{(1-s)Q(s)}},
\ene
listed in Table~\ref{tab:summary}. Since $R(-\chi)=(1+\chi)^2(2-\chi)/3>0$,
the column $\Sigma_s=\sqrt{R(-\chi)}$ remains finite there, and so does the
Newtonian field. Integrating
$d[(\rho+\chi)^2]/dx\simeq-2\Sigma_s$ near the surface gives the
square-root profile
\bee
\rho\simeq-\chi+\sqrt{2\Sigma_s\,(x_s-|x|)} .
\label{eq:sqrt}
\ene
The density gradient and the field $g=-\rho\Sigma/(\rho+\chi)$ diverge
integrably. At the surface the effective sound speed and the local Jeans
length vanish. Beyond it $\ceff^2<0$, and the medium has negative
effective compressibility. A regular hydrostatic continuation to lower
densities is therefore impossible. The termination is a genuine
mechanical limit of the effective fluid, not an artifact of the
integration. How such a surface would be completed, for example by a
density jump or by physics beyond the Newtonian EiBI limit, is outside the
scope of this paper. The effective pressure is negative at the surface,
$\peff(-\chi)=\chi^3/6<0$, which shows that the surface cannot be closed by
simple pressure balance with a vacuum.

For $\chi=-2/3$ we have $a=0$ and $Q=\rho^2$, and Eq.~(\ref{eq:adia_quad})
is elementary:
\bee
|x|=\sqrt6\left[\sqrt{1-\rho}-\frac23\,\artanh\sqrt{1-\rho}\right],
\qquad \frac23\le\rho\le1 ,
\label{eq:chim23}
\ene
with $x_s=\sqrt6\,[3^{-1/2}-\tfrac23\artanh(3^{-1/2})]\simeq0.339$. The
elementary cases $\chi=2$ and $\chi=-2/3$ are exactly the two parameter
values at which $Q$ has a double root.

\begin{figure}[t]
\centering
\includegraphics[width=\linewidth]{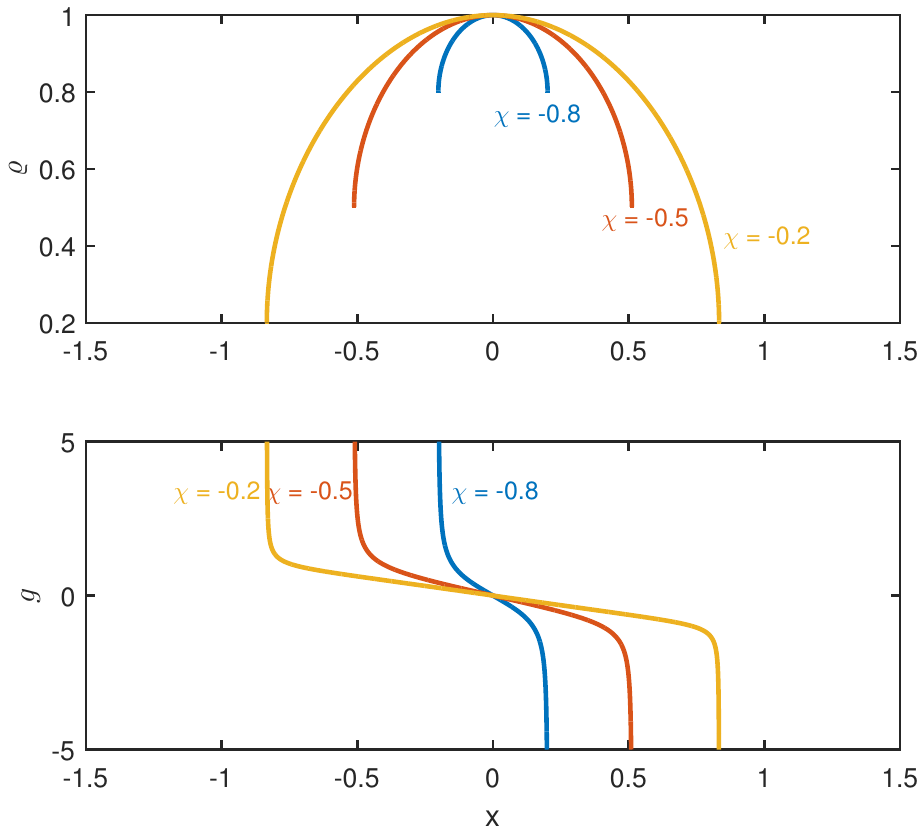}
\caption{Adiabatic ($\gamma=3$) sheets for $\chi=-0.8$, $-0.5$ and $-0.2$:
density (top) and field $g=\rho\rho'$ (bottom). Each profile ends at the
finite density $\rho=-\chi$, where the effective sound speed
$\ceff^2=\rho(\rho+\chi)$ vanishes. The density approaches this value as
the square root of Eq.~(\ref{eq:sqrt}), and $g$ diverges. The sheet
thickness $x_s$ decreases as $\chi\to-1$ (Table~\ref{tab:summary}).}
\label{fig:neg_chi_profiles}
\end{figure}

\begin{table}[t]
\centering
\caption{Characteristic quantities of the planar equilibria (dimensionless
units). $x_{1/2}$: half-width at half-maximum; $\Sigma_b$: half-column;
$x_b$: position of the boundary (edge $x_e$ for $\chi\ge0$, singular
surface $x_s$ for $\chi<0$) of the $\gamma=3$ sheet; $\rho_b$: boundary
density. Isothermal sheets are unbounded, with $\Sigma_b=\sqrt{2+\chi}$.
All values were obtained by quadrature of Eqs.~(\ref{eq:iso_quad}) and
(\ref{eq:adia_quad}).}
\label{tab:summary}
\begin{tabular}{rccccc}
\toprule
 & \multicolumn{2}{c}{Isothermal} & \multicolumn{3}{c}{Adiabatic, $\gamma=3$}\\
\cmidrule(lr){2-3}\cmidrule(lr){4-6}
$\chi$ & $x_{1/2}$ & $\Sigma_b$ & $\rho_b$ & $x_b$ & $\Sigma_b$\\
\midrule
$-0.8$ & 0.828 & 1.095 & 0.800 & 0.202 & 0.193\\
$-2/3$ & 0.910 & 1.155 & 0.667 & 0.339 & 0.314\\
$-0.5$ & 1.004 & 1.225 & 0.500 & 0.513 & 0.456\\
$-0.2$ & 1.156 & 1.342 & 0.200 & 0.835 & 0.685\\
$0$    & 1.246 & 1.414 & 0     & 1.056 & 0.816\\
$0.5$  & 1.449 & 1.581 & 0     & 1.524 & 1.080\\
$1$    & 1.627 & 1.732 & 0     & 1.882 & 1.291\\
$2$    & 1.935 & 2.000 & 0     & 2.449 & 1.633\\
$5$    & 2.652 & 2.646 & 0     & 3.660 & 2.380\\
\bottomrule
\end{tabular}
\end{table}

\begin{figure}[t]
\centering
\includegraphics[width=\linewidth]{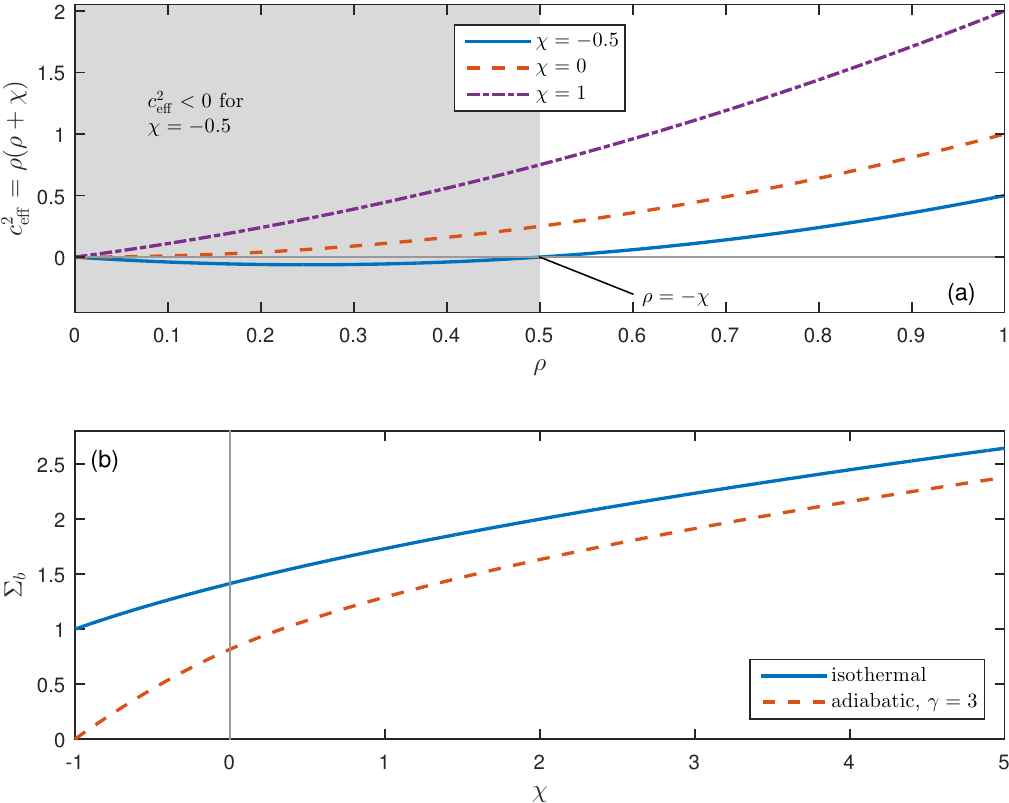}
\caption{(a) Effective sound speed $\ceff^2=\rho(\rho+\chi)$ of the
$\gamma=3$ fluid for $\chi=-0.5$, $0$ and $1$. For $\chi<0$ it vanishes at
$\rho=-\chi$ and is negative below (shaded for $\chi=-0.5$), which is
where the equilibria of Fig.~\ref{fig:neg_chi_profiles} terminate. (b)
Half-column $\Sigma_b$ against $\chi$ for isothermal sheets
[$\sqrt{2+\chi}$] and $\gamma=3$ sheets [Eq.~(\ref{eq:Sigmab}), or
$\sqrt{R(-\chi)}$ for $\chi<0$].}
\label{fig:ceff_sigma}
\end{figure}

Figure~\ref{fig:ceff_sigma} summarizes the role of the effective sound
speed and the dependence of the column on $\chi$ for both equations of
state. In the adiabatic case $\Sigma_b\to0$ as $\chi\to-1^+$, because the
singular surface moves to the midplane. In the isothermal case
$\Sigma_b\to1$, which is the maximum column of Eq.~(\ref{eq:Sigmamax}) in
dimensionless form.

\FloatBarrier

\section{Bounded isothermal spheres and the EiBI Bonnor--Ebert mass}
\label{sec:sphere}
The planar first integral (\ref{eq:firstint}) has no spherical analogue,
but the effective-pressure reduction carries over unchanged and, as we now
show, extends to the dynamics.

\subsection{Exact Newtonian form of the dynamics}
Equation~(\ref{eq:poisson_intro}) contains no time derivative, so at every
instant $\psi=\psi_N+\chi\rho/4$, with $\nabla^2\psi_N=4\pi G\rho$ (the
difference is harmonic and vanishes at infinity). Hence
$\rho\nabla\psi=\rho\nabla\psi_N+\nabla(\chi\rho^2/8)$, and the momentum
equation in Eq.~(\ref{eq:fluid}) becomes
\bee
\rho\left(\fpar{\mathbf{v}}{t}+\mathbf{v}\cdot\nabla\mathbf{v}\right)
=-\nabla\peff-\rho\nabla\psi_N
\label{eq:euler_eff}
\ene
for any flow, not only in equilibrium. The Newtonian EiBI fluid is
therefore exactly a Newtonian self-gravitating fluid with pressure
$\peff$. For perturbations that preserve the isothermal law, $\peff(\rho)$
is barotropic, and the classical equilibrium and stability theory of
Newtonian configurations applies with $c_s^2$ replaced by
$\ceff^2=c_s^2+\chi\rho/4$.

\subsection{Boundary condition and parameters}
Following \citet{Bonnor1956} and \citet{Ebert1955}, we consider an
isothermal sphere of radius $R$ embedded in a tenuous medium of pressure
$P_{\rm ext}$, whose density is small enough that the EiBI term is
negligible outside. Integrating Eq.~(\ref{eq:euler_eff}) across the
density discontinuity at $r=R$, where $\psi_N'$ is continuous, shows that
$\peff$ is continuous:
\bee
c_s^2\rho_R+\frac{\chi}{8}\rho_R^2=P_{\rm ext}.
\label{eq:match}
\ene
The thermal pressure alone is not continuous. The term
$-\rho\chi\nabla\rho/4$ in the EiBI force is concentrated at the surface
and supplies the jump $\chi\rho_R^2/8$; this is the spherical counterpart
of the edge-field discontinuity of the planar sheets in
Sec.~\ref{sec:adia}. A condition $c_s^2\rho_R=P_{\rm ext}$ would not
conserve momentum across the surface.

In the units of Eq.~(\ref{eq:norm}) with $\rho_0\to\rho_c$, the central
density, the equilibrium satisfies the modified Emden equation
$r^{-2}(r^2h')'=-\rho$ with $h=\ln\rho+\hat\chi\rho$, or equivalently
\bee
\fder{\rho}{r}=-\frac{m\,\rho}{r^2(1+\hat\chi\rho)},\qquad
\fder{m}{r}=r^2\rho ,
\label{eq:sph_struct}
\ene
with $\rho(0)=1$. Near the centre
$\rho\simeq1-r^2/[6(1+\hat\chi)]$, so a regular centre again requires
$\hat\chi>-1$. As $\hat\chi\to-1^+$ the profile develops a cusp: with
$h\simeq-1-(1-\rho)^2/2$ and $\nabla^2h\simeq-1$ one finds
$\rho\simeq1-r/\sqrt3$, and the mass stays finite.

For a physical cloud, $\chi$, $c_s$ and $P_{\rm ext}$ are fixed, while the
central density varies along the sequence of equilibria. We therefore
define $\rho_P=P_{\rm ext}/c_s^2$ and the single EiBI parameter
\bee
\varepsilon=\frac{\chi P_{\rm ext}}{4c_s^4}=\frac{\chi\rho_P}{4c_s^2},
\label{eq:eps}
\ene
which is the value $\hat\chi$ would take at density $\rho_P$;
$\varepsilon/2$ is the ratio of the EiBI to the thermal pressure at that
density. Each equilibrium is labelled by $D=\rho_c/\rho_P$, with
$\hat\chi=\varepsilon D$. The boundary $\xi_R$ is where
$\rho+\hat\chi\rho^2/2=1/D$, and the mass in Bonnor--Ebert units is
\bee
\mu\equiv\frac{M\,G^{3/2}P_{\rm ext}^{1/2}}{c_s^4}
=\frac{m(\xi_R)}{\sqrt{4\pi D}} .
\label{eq:mu}
\ene
For $\varepsilon=0$ we recover the classical maximum $\mu_{\rm
BE}=1.1822$ at $D=14.042$, $\xi_R=6.451$. For $\varepsilon<0$ the
regularity bound restricts the sequence to $D<1/|\varepsilon|$: the
central density cannot exceed $4c_s^2/|\chi|$, the same density that
bounds the planar sheets in Eq.~(\ref{eq:Sigmamax}).

\subsection{Radial stability}
\begin{figure}[t]
\centering
\includegraphics[width=\textwidth]{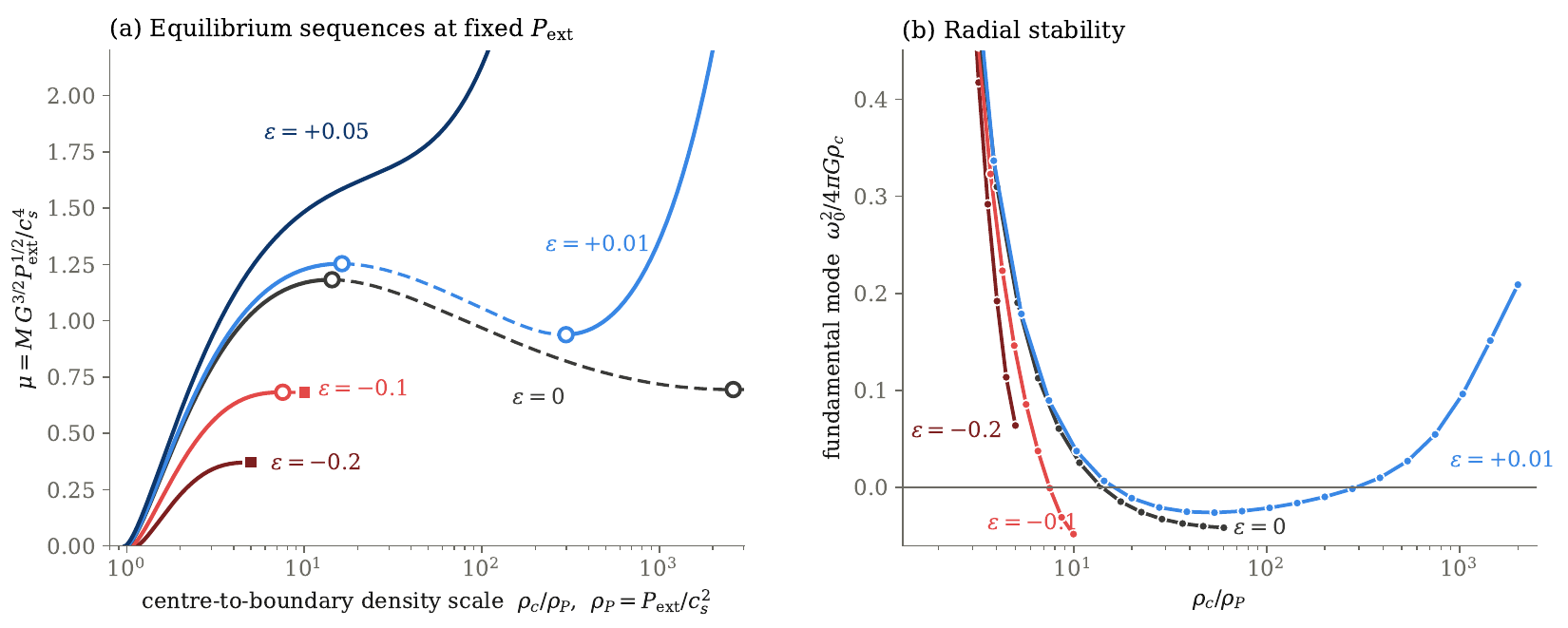}
\caption{(a) Sequences of pressure-bounded isothermal spheres at fixed
$\chi$, $c_s$ and $P_{\rm ext}$, labelled by
$\varepsilon=\chi P_{\rm ext}/4c_s^4$: dimensionless mass $\mu$ against
central density $\rho_c/\rho_P$. Solid curves are stable and dashed curves
unstable; open circles mark mass extrema and filled squares the end of
the $\varepsilon<0$ sequences at $\hat\chi=-1$. (b) Fundamental radial
eigenvalue $\omega_0^2$ along the same sequences. It vanishes at each mass
extremum; for $\varepsilon=0.01$ it becomes positive again at the mass
minimum, and for $\varepsilon=-0.2$ it stays positive up to the end of the
sequence.}
\label{fig:sph_seq}
\end{figure}

Because of Eq.~(\ref{eq:euler_eff}), radial isothermal perturbations
obey the Newtonian linear adiabatic wave equation with
$\Gamma=\rho\ceff^2/\peff$. Writing $y_1=\xi_r/r$ and
$y_2=\Delta\peff/\peff$ for the Lagrangian displacement and pressure
perturbation, with time dependence $e^{i\omega t}$ and $\omega^2$ in units
of $4\pi G\rho_c$,
\beea
r\fder{y_1}{r}&=&-3y_1-\frac{y_2}{\Gamma},\non\\
\fder{y_2}{r}&=&\frac{\rho}{\peff}\left[\frac{m}{r^2}\,y_2
+\left(\omega^2r+\frac{4m}{r^2}\right)y_1\right],
\label{eq:lawe}
\enea
with $y_2=-3\Gamma y_1$ at the centre and $y_2(\xi_R)=0$, i.e. a fixed
external effective pressure. We solved Eq.~(\ref{eq:lawe}) by shooting
for the fundamental eigenvalue $\omega_0^2$ and checked the results
independently with a Galerkin (Rayleigh--Ritz) discretization of the same
problem; the two agree to at least four significant figures.
Figure~\ref{fig:sph_seq}(b) shows that $\omega_0^2$ changes sign exactly
at the extrema of $\mu(D)$, as the turning-point
method \citep{Katz1978} requires, and nowhere else.

\subsection{Three regimes of the critical mass}
\begin{figure}[t]
\centering
\includegraphics[width=\textwidth]{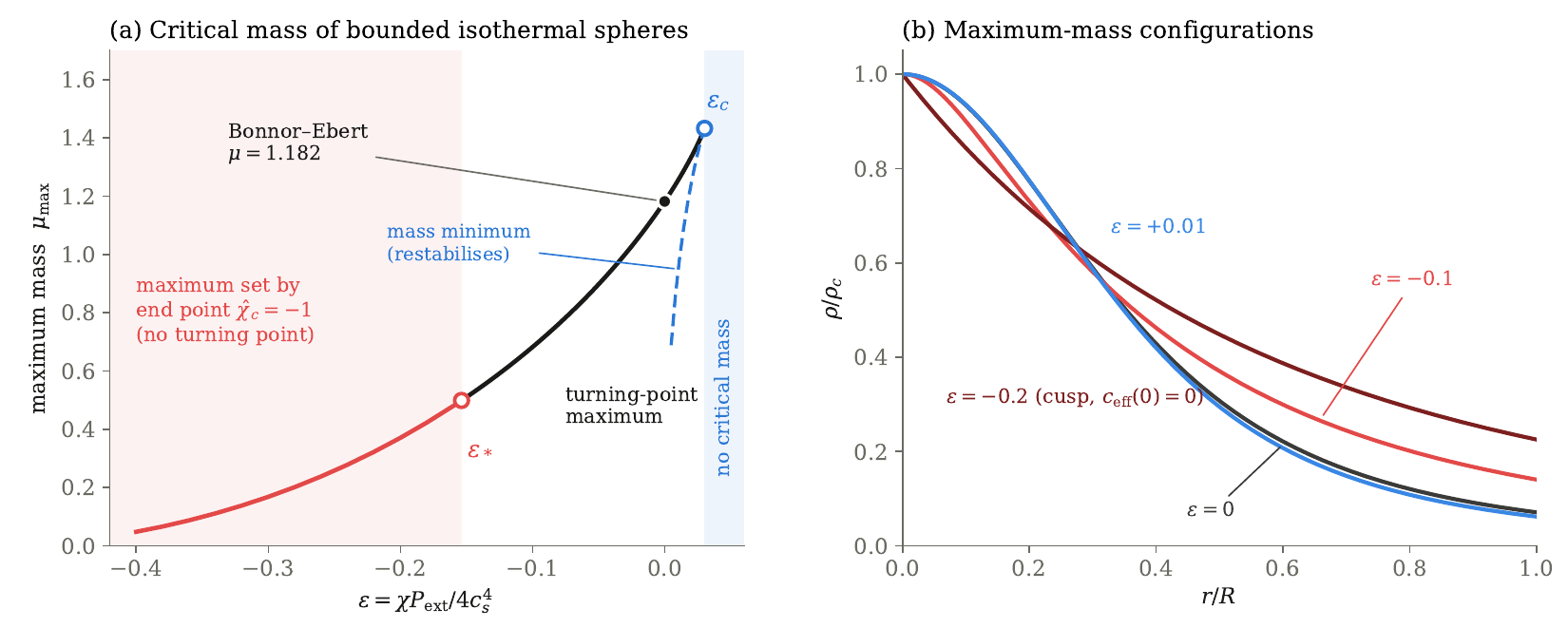}
\caption{(a) Maximum mass $\mu_{\max}$ of pressure-bounded isothermal
spheres against $\varepsilon$. Black: turning-point maximum. Red: end-point
maximum at $\hat\chi=-1$ for $\varepsilon<\varepsilon_*$. Dashed blue: the
mass minimum at which the fundamental mode restabilizes for
$0<\varepsilon<\varepsilon_c$. For $\varepsilon\ge\varepsilon_c$ there is
no critical mass. (b) Density profiles of the maximum-mass
configurations; the $\varepsilon=-0.2$ profile is cusped,
$\rho\simeq\rho_c(1-r/\sqrt3\lambda_J)$, because $\ceff$ vanishes at the
centre.}
\label{fig:sph_crit}
\end{figure}

Figure~\ref{fig:sph_seq}(a) shows representative sequences, and
Fig.~\ref{fig:sph_crit}(a) and Table~\ref{tab:sphere} give the maximum
mass as a function of $\varepsilon$. Three regimes appear.

\emph{(i) Shifted Bonnor--Ebert turning point,
$\varepsilon_*<\varepsilon<\varepsilon_c$.} The mass has an interior
maximum at which the fundamental mode becomes unstable. The critical mass
increases monotonically with $\varepsilon$; near $\varepsilon=0$,
\bee
M_{\rm crit}\simeq1.182\,(1+5.74\,\varepsilon)\,
\frac{c_s^4}{G^{3/2}P_{\rm ext}^{1/2}} .
\label{eq:Mcrit_lin}
\ene
A positive $\chi$ raises the critical mass and a negative $\chi$ lowers
it, consistent with the trends of the homogeneous Jeans
analyses \citep{Yang2023,Hazarika2024}, but here quantified for the
bounded, stratified configuration. For $0<\varepsilon<\varepsilon_c$ the
unstable branch ends at a mass minimum, where the fundamental mode
restabilizes [Fig.~\ref{fig:sph_seq}(b)], and beyond it the mass grows
without bound. On this high-density branch $\hat\chi\rho\gg1$ in the core,
which is supported by the EiBI pressure as an $n=1$ polytrope. For a
cloud whose mass exceeds the critical value, a stable equilibrium of the
same mass therefore exists at much higher central density,
$\rho_c\gtrsim4c_s^2/\chi$, instead of the singular collapse of the
Newtonian case. Whether a collapsing cloud actually settles on this
branch is a dynamical question that we do not address here.

\emph{(ii) No critical mass, $\varepsilon\ge\varepsilon_c$.} The maximum
and the minimum approach each other as $\varepsilon$ increases and merge
at
\bee
\varepsilon_c=0.0302,\qquad \hat\chi\simeq1.05,\quad D\simeq35,\quad
\mu\simeq1.433 .
\ene
For $\varepsilon>\varepsilon_c$, $\mu(D)$ increases monotonically, and
$\omega_0^2>0$ along the entire sequence. Every pressure-bounded
isothermal sphere is then radially stable, whatever its mass: a positive
EiBI parameter with $\chi P_{\rm ext}/c_s^4>4\varepsilon_c\simeq0.121$
removes the Bonnor--Ebert instability altogether.

\emph{(iii) Mass limit without instability,
$\varepsilon<\varepsilon_*$.} As $\varepsilon$ decreases, the turning
point moves to higher $|\hat\chi|$ and reaches the end of the sequence,
$\hat\chi=-1$, at
\bee
\varepsilon_*=-0.1537,\qquad \mu=0.499 .
\ene
For $\varepsilon<\varepsilon_*$, $\mu(D)$ increases monotonically up to
the end point, and $\omega_0^2$ remains positive there (for example
$\omega_0^2=0.064$ at $\varepsilon=-0.2$). The maximum mass is the
end-point value, approached as $\hat\chi\to-1$, where the central
effective sound speed vanishes and the profile becomes cusped
[Fig.~\ref{fig:sph_crit}(b)]. A cloud of larger mass has no equilibrium,
but the configurations that do exist are not unstable: the limit is a
loss of equilibrium, not a dynamical instability. This is the spherical
counterpart of the maximum column of planar sheets. In the planar case
Eq.~(\ref{eq:firstint}) gives $\peff(\rho_0)-P_{\rm ext}=\pi
G\Sigma_{\rm tot}^2/2$ for a bounded sheet, which increases with $\rho_0$
until $\ceff^2(\rho_0)=0$, so the end point is the only limit. Spherical
geometry adds the gravitational turning point, which takes over for
$\varepsilon>\varepsilon_*$.

\begin{table}[t]
\caption{Maximum-mass configurations of pressure-bounded isothermal
spheres. $D=\rho_c/\rho_P$; $\hat\chi=\varepsilon D$; $\xi_R$: radius in
units of $\lambda_J(\rho_c)$; $\mu_{\max}$ in units of
$c_s^4/G^{3/2}P_{\rm ext}^{1/2}$. ``End'' denotes the end-point maximum
at $\hat\chi\to-1$. For $\varepsilon>0$ the last two columns give the
mass minimum at which the fundamental mode restabilizes.}
\label{tab:sphere}
\begin{tabular}{rcrrrrrr}
\toprule
$\varepsilon$ & type & $D$ & $\hat\chi$ & $\xi_R$ & $\mu_{\max}$ &
$D_{\min}$ & $\mu_{\min}$\\
\midrule
$-0.30$ & end     & 3.333  & $-1$     & 1.904 & 0.1686 & & \\
$-0.20$ & end     & 5.000  & $-1$     & 2.992 & 0.3710 & & \\
$-0.15$ & turning & 6.456  & $-0.968$ & 3.723 & 0.5106 & & \\
$-0.10$ & turning & 7.514  & $-0.751$ & 4.326 & 0.6825 & & \\
$-0.05$ & turning & 9.491  & $-0.475$ & 5.149 & 0.8977 & & \\
$-0.02$ & turning & 11.603 & $-0.232$ & 5.827 & 1.0569 & & \\
$0$     & turning & 14.042 & $0$      & 6.451 & 1.1822 & & \\
$0.01$  & turning & 16.029 & $0.160$  & 6.874 & 1.2535 & 289.6 & 0.9379\\
$0.02$  & turning & 19.386 & $0.388$  & 7.471 & 1.3335 & 105.9 & 1.2421\\
$0.03$  & turning & 32.053 & $0.962$  & 9.010 & 1.4305 & 39.6  & 1.4303\\
\bottomrule
\end{tabular}
\end{table}
\FloatBarrier

\section{Discussion}
\label{sec:disc}
\emph{Relation to earlier work.} The effective-pressure form of the
Newtonian EiBI equations is known for hydrostatic
equilibrium \citep{Pani2011,Pani2012} and underlies the modified
Lane--Emden equations used for EiBI and related Ricci-based
theories \citep{Olmo2020,Wibisono2018,Orazi2020}. Two points here are new.
First, Eq.~(\ref{eq:euler_eff}) shows that the equivalence holds for the
full dynamics, which justifies applying Newtonian stability theory with
$\peff$ and fixes the boundary condition (\ref{eq:match}). Second, it
yields the critical mass as a function of the single parameter
$\varepsilon$, including the two qualitative changes at
$\varepsilon_c$ and $\varepsilon_*$, neither of which is visible in a
perturbation analysis about a uniform medium. \citet{PaniSotiriou2012} showed that EiBI polytropic stars can
develop surface curvature singularities because the field depends on
density gradients. The planar $\gamma=3$ sheets display both faces of
this problem in exactly solvable form: for $\chi>0$ the field acting on
the fluid is discontinuous at the edge, and for $\chi<0$ the density
gradient diverges at the $\ceff^2=0$ surface.

\emph{Relation to Jeans-instability studies.} The linear and nonlinear
EiBI Jeans analyses of
\citet{DeMartino2017,Bessiri2021,Yang2023,Hazarika2024,DasKarmakar2024Puls,DasKarmakar2024NL}
identify $\ceff^2=c_s^2+\chi\rho_0/4$ as the quantity that controls the
stability of a uniform medium. In the bounded configurations studied
here the same quantity, evaluated locally, decides more: a regular centre
requires $\ceff^2(\rho_c)>0$, $\gamma=3$ sheets terminate where $\ceff^2$
vanishes, and for $\varepsilon<\varepsilon_*$ the vanishing of the central
$\ceff$ replaces the Bonnor--Ebert instability as the mass limit.

\emph{Magnitude of $\chi$ and other interpretations.} The parameter
$\varepsilon$ compares the EiBI and thermal pressures at the boundary
density. Current bounds on $\chi$ \citep{Avelino2012} imply that
$\varepsilon$ is negligible for molecular clouds, for which
Eq.~(\ref{eq:Mcrit_lin}) shows that the Bonnor--Ebert mass is unchanged
in practice, and that it becomes appreciable only in dense
matter \citep{DeMartino2017}. The same equations describe, with
$\chi/8$ replaced by $2\pi\hbar^2a_s/m^3$, a self-gravitating fluid whose
equation of state contains the contact-interaction term of a
Bose--Einstein condensate with scattering length
$a_s$ \citep{Boehmer2007,Chavanis2011}. For attractive interactions
($a_s<0$) condensates are known to have a maximum
mass \citep{Chavanis2016}; our results give its counterpart for a
thermally supported, pressure-bounded cloud, including the change from a
turning-point to an end-point limit at $\varepsilon_*$. For repulsive
interactions they show that the isothermal Bonnor--Ebert instability is
suppressed when $\varepsilon>\varepsilon_c$.

\emph{Extensions.} Natural next steps are non-radial and transverse
perturbations of the planar sheets \citep{Ledoux1951}, cylindrical
filaments, and the nonlinear collapse of supercritical clouds, which
would show whether, for $0<\varepsilon<\varepsilon_c$, collapse ends on
the stable EiBI-supported branch.

\section{Conclusions}
\label{sec:concl}
Treating the EiBI correction as an effective pressure
$\peff=p+\chi\rho^2/8$, which we have shown is exact for arbitrary
Newtonian flows, reduces self-gravitating EiBI fluids to Newtonian ones
with a modified equation of state. The principal results are:
\begin{enumerate}
\item For pressure-bounded isothermal spheres, the critical mass depends
on the single parameter $\varepsilon=\chi P_{\rm ext}/4c_s^4$. It grows
as $M_{\rm BE}(1+5.74\varepsilon)$ near $\varepsilon=0$, and the
turning-point instability disappears altogether for
$\varepsilon>\varepsilon_c\simeq0.0302$. For
$0<\varepsilon<\varepsilon_c$ a stable high-density branch supported by
the EiBI pressure follows the unstable one.
\item For $\varepsilon<\varepsilon_*\simeq-0.154$ the mass limit is set by
the vanishing of the central effective sound speed at
$\chi\rho_c/4c_s^2=-1$; it is a loss of equilibrium, not an instability.
\item In planar geometry the first integral
$\peff(0)-\peff(x)=2\pi G\Sigma^2$ gives closed-form isothermal sheets for
all $\hat\chi>-1$, with a maximum column density
$\Sigma_{\max}=2c_s^2/\sqrt{\pi G|\chi|}$ for $\chi<0$ that is the planar
counterpart of the end-point limit.
\item Adiabatic $\gamma=3$ sheets with $-1<\chi<0$ always terminate at
$\rho=-\chi$, where the effective sound speed vanishes. For $\chi>0$ they
have compact support with a finite edge slope, and the cases $\chi=2$
(the parabola $\rho=1-x^2/6$) and $\chi=-2/3$ are elementary.
\end{enumerate}
All results reduce to the classical Spitzer sheets and Bonnor--Ebert
spheres as $\chi\to0$.

\section*{Statements and Declarations}

\paragraph{Funding} No funding was received for conducting this study.

\paragraph{Competing interests} The author has no relevant financial or
non-financial interests to disclose.

\paragraph{Author contributions} The author is the sole contributor to
the conception, analysis, computations and writing of this work.

\paragraph{Data availability} The data that support the findings of this
study are available within the article.

\paragraph{Code availability} The numerical codes used for
Sec.~\ref{sec:sphere} (Python and MATLAB versions) are available from the
author on reasonable request.

\bibliographystyle{plainnat}
\bibliography{eibiref_arxiv}
\end{document}